\documentclass{ai4x}

\usepackage{fancyhdr}
\usepackage{subcaption}

\IACpaperyear{2025} 
\IAClocation{Singapore} 
\IACdate{8--11 July 2025} 

\title{Conversational Self-Play for Discovering and Understanding Psychotherapy Approaches}

\hypersetup{
    pdfkeywords={self-play, LLM, AI, psychotherapy, AI for Science, digital mental health}
}

\begin{document}

\twocolumn[
\maketitle
\icmlsetsymbol{equal}{*}
\begin{icmlauthorlist}
\icmlauthor{Onno P. Kampman}{moht}{0000-0001-8125-2565}{presenting}
\icmlauthor{Michael Xing}{moht}{}{}
\icmlauthor{Charmaine Lim}{moht}{0000-0002-6006-6138}{}
\icmlauthor{Ahmad Ishqi Jabir}{fht}{0000-0001-6864-9867}{}
\icmlauthor{Ryan Louie}{stanford}{}{}
\icmlauthor{Jimmy Lee}{imh}{0000-0002-7724-7445}{}
\icmlauthor{Robert JT Morris}{moht}{0000-0001-8320-2393}{}
\end{icmlauthorlist}

\icmlaffiliation{moht}{MOH Office for Healthcare Transformation, Singapore}
\icmlaffiliation{fht}{Future Health Technologies, Singapore-ETH Centre, Campus for Research Excellence And Technological Enterprise, Singapore}
\icmlaffiliation{stanford}{Stanford University, USA}
\icmlaffiliation{imh}{Institute of Mental Health, Singapore}
\icmlcorrespondingauthor{}{onno.kampman@moht.com.sg}

\printAffiliations
\vskip 2.5ex
]
\thispagestyle{fancy} 

\section{Introduction}
\label{sec:introduction}

While artificial intelligence (AI) has accelerated scientific discovery across domains such as drug discovery, protein folding~\cite{jumperHighlyAccurateProtein2021}, and materials science~\cite{gottweisAICoscientist2025}, its application to understanding effective psychotherapy remains limited.
Although large language models (LLMs) are increasingly deployed to analyze, assist, and replace human therapists in psychotherapeutic conversations~\cite{hua2024large, kampman2024ai, sharma2023human, siddals2024it}, these applications primarily replicate established therapeutic approaches such as Cognitive Behavioral Therapy (CBT) and Motivational Interviewing (MI).
They do not address fundamental research questions regarding when and for whom specific therapeutic approaches are effective, nor do they illuminate the underlying mechanisms of therapeutic success.

Traditional psychotherapy research methods, including randomized control trials and longitudinal studies, have yielded valuable insights into treatment effectiveness.
However, these approaches are constrained by their time-intensive nature, resource demands, and limited capacity to systematically explore the vast combinatorial space of therapeutic approaches and their adaptations to diverse patient presentations~\cite{imelComputationalPsychotherapyResearch2015, markowitz_lost_2022, groth_psychotherapy_2021}.

We propose that LLMs offer a scalable and systematic \emph{platform} to assess, discover, and improve therapeutic approach selection by extracting latent knowledge encoded within these high-performance systems.
This computational framework enables systematic exploration of emergent or effective psychotherapeutic processes through conversational simulations, implemented via \emph{self-play}--a reinforcement learning (RL) technique where agents interact with copies of themselves.
The self-play mechanism that enabled AlphaGo to develop self-directed strategies in the game of Go \emph{without} human guidance, surpassing human expertise and revealing previously unknown tactical approaches~\cite{silver2016mastering}, exemplifies this potential.
The famous ``Move 37'' surprised human players, who found it creative and initially hard to grasp, and subsequently inspired work to analyze its effectiveness~\cite{schutBridgingHumanAIKnowledge2023a}.

Analogously, we simulate turn-based therapeutic dialogues~\cite{li2024mentalarena}, even though therapy is collaborative rather than competitive~\cite{bardHanabiChallengeNew2020, liaoEfficacyLanguageModel2024}.
Two LLM agents assume the roles of therapist and client~\cite{shanahan2023role, parkGenerativeAgentsInteractive2023}, operating within the constraints of single-session, text-based therapy--a format commonly found in anonymous helplines~\cite{kim2023effectiveness}.
This controlled environment isolates the impact of therapeutic approach selection and techniques used (such as cognitive restructuring from CBT or confidence ruler from MI) from factors including therapeutic alliance, non-verbal cues, and longitudinal relationship development.

Emergent conversational patterns~\cite{lazaridouEmergentMultiAgentCommunication2020} reveal the techniques enacted by AI therapists and their alignment with established therapeutic approaches.
Of particular interest are \emph{deviations} from standard approaches, including novel therapeutic techniques, innovative sequencing of techniques within sessions, unconventional applications of established, and adaptive approaches tailored to client characteristics.
This paper presents a proof-of-concept study demonstrating how AI can serve as a discovery engine for psychotherapy research.

\section{Methods}
\label{sec:methods}

We simulated therapeutic dialogues between two LLMs agents (\emph{actors}) (see Tab.~\ref{tab:example-conversation}).
One agent role-played as therapist, while the other enacted a client presenting with depressive symptoms at five severity levels based on DSM-5 criteria~\cite{APA2013DSM5} (see Tabs.~\ref{tab:dsm-criteria} and~\ref{tab:actor-descriptions}).
Actor behaviors were differentiated through detailed system prompts or \emph{vignettes}~\cite{louie2024roleplay, park2024generative, priyadarshana2024prompt}, developed and validated by a counselor who confirmed the resulting transcripts achieved sufficient clinical realism.
Each client persona was assigned a comprehensive background, context, emotional struggles, objective for therapy, and personality matched to severity level (see Tab.~\ref{tab:actor-system-prompt}).
This stratification was designed to elicit distinct therapeutic approaches, as clients with severe depression typically exhibit greater functional impairment.
The AI therapist (see Tab.~\ref{tab:therapist-system-prompt}) was initialized with counseling microskills (e.g., active listening, empathy), but was not restricted to any therapeutic approach.
This design enabled observation of spontaneous therapeutic approach adoption and integration based on client presentations, rather than adherence to predetermined protocols.

We generated 100 self-play sessions per client persona using GPT-4o~\cite{openai2024gpt4o}, with each dialogue comprising 10 conversational turns per agent.
A separate LLM agent, functioning as a research analyst, evaluated the therapeutic approaches (see Tab.~\ref{tab:therapeutic-approaches}) and specific techniques (e.g., asking scaling questions, resolving ambivalence; see Tab.~\ref{tab:therapeutic-techniques}) employed by the AI therapist.

\begin{figure}[t]
    \centering
    \includegraphics[width=\columnwidth]{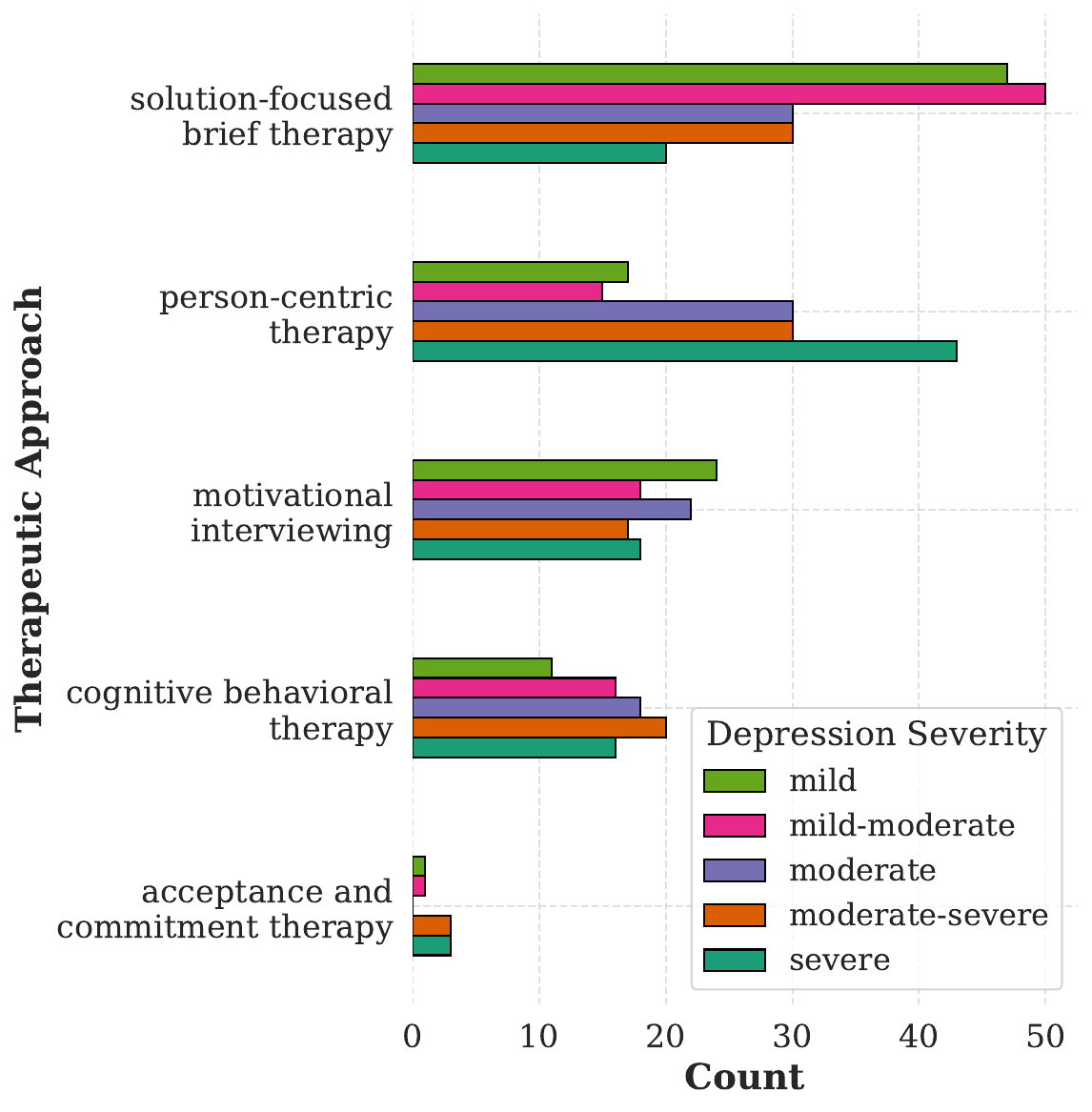}
    \caption{Therapeutic approaches used by AI therapist.}
    \label{fig:results-approaches}
\end{figure}

\section{Results and Discussion}
\label{sec:results-discussion}

The AI therapist predominantly employed Solution-Focused Brief Therapy (SFBT), Person-Centered Therapy (PCT), CBT, and MI approaches--all well-suited for brief, text-based interactions and commonly used in counseling sessions with an individual with depression (Fig.~\ref{fig:results-approaches}).
The preponderance of SFBT likely reflects mainstream LLM-human alignment with constructive, solution-oriented responses that avoid deeper psychological exploration.
Notably, SFBT usage decreased with symptom severity while PCT increased, suggesting adaptive selection based on client characteristics.
Milder cases may benefit from a more straightforward application of SFBT techniques like exception finding and goal setting, whereas severe cases require PCT's open-ended questioning and reflective listening for deeper exploration.
Open-ended questioning, reflective listening, and behavioral activation were the most prevalent individual techniques (Fig.~\ref{fig:results-techniques-combined}).
These findings demonstrate that self-play can uncover patterns in LLM-generated therapeutic conversations that replicate known dynamics in human therapy.

Several limitations warrant future investigation.
First, general-purpose LLMs may inadequately simulate realistic therapeutic actors.
Domain-specific, fine-tuned models could enhance simulation fidelity~\cite{stade2024large, xu2024mental, li2024mentalarena}, requiring continued attention to validation and alignment with real client presentations~\cite{wangClientCenteredAssessmentLLM2024}.
Second, while we draw analogies to RL, which often involves training models to refine a policy based on rewards, our approach did not involve RL fine-tuning.
Instead, we treated the AI therapist's style and approach as implicit policy--a mapping from client states (\emph{situations}) to therapist actions (\emph{responses}), which can be explicitly prespecified (rule-based) or learned from experience--to explore the capabilities and policies already present in the model.
If we want to evaluate effectiveness of conversations or improve model policies through RL self-play, further research is warranted to define clear reward functions, such as whether a conversation was helpful or not~\cite{althoff2016large}.
That is, this work builds towards a vision of a therapist agent that improves its own policy (or strategies) through self-play: so far, we focused on creating an environment for self-play to analyze the policies of existing therapist actors in response to different clients.
This extension provides the foundations for studying therapy through simulacra of generative agents~\cite{parkGenerativeAgentsInteractive2023}.
Third, our framework oversimplifies psychotherapy by assuming single-approach sessions.
Like a Go grandmaster or professional athlete, effective therapists adaptively draw from multiple approaches \emph{during} conversations based on what the client says or needs and the state or stage of the conversation, with the final goal of building therapeutic alliance and helping them reach their goals.
Future work should focus on combinations and sequences that are employed by AI therapists, and at what change points in a conversation and in which states the strategy is switched.
Fourth, this platform enables investigation of treatment ``core components'' versus adaptable elements of treatment approaches which can be changed without compromising the efficacy of the treatment~\cite{goldsmith_evaluation_2023}.
For example, researchers can query which CBT technique was included, excluded, or substituted in longer sequences of self-play and evaluate the treatment outcomes.
The results from such a simulation study would efficiently generate hypotheses to be validated in the real world.
Finally, while designed to discover \emph{emergent} techniques, this proof-of-concept primarily replicated established approaches through classification analysis.
Future work should employ clustering, outlier analysis, and (human-rated) qualitative methods to analyze therapeutic approaches and identify novel patterns, distinguishing truly innovative techniques from underutilized existing ones.
Even seemingly counterproductive approaches--such as confrontation in therapy--often trace to established frameworks like reactive emotive behavior therapy, making novelty definition challenging.
This platform should ultimately study both novel therapeutic patterns and fundamental, adaptable components of effective interventions while ensuring evidence-based, safe therapeutic practices.

\clearpage

\section*{Acknowledgments}

This study exclusively used simulated data without human participants or evaluators, thus institutional review board approval was not required.
The use of commercial APIs was deemed appropriate given the absence of human subject involvement.

Since this study involves AI-generated simulations rather than real patients, ethical concerns primarily revolve around model biases and the risk of overgeneralizing AI-driven insights to real-world clinical practice.
All findings are presented as exploratory with explicit acknowledgement of LLM limitations and the critical need for human oversight in AI-assisted mental health applications.

\section*{Data and Software Access}

Simulated conversational transcripts are available upon request.

Access to our UI render\footnote{System is hosted at \url{https://dialogue.ai.mohtgroup.com}} of simulated self-play conversations--shown in Fig.~\ref{fig:self-play-example}--is also available on request.

\bibliography{biblio}

\clearpage

\onecolumn

\begin{appendices}

\section{Background Details}

\begin{table}[htp]
    \centering
    \caption{
        DSM-5 criteria for major depressive episode.
        For a clinical diagnosis of major depressive disorder (MDD), symptoms should be present nearly every day during the same 2-week period, and low mood or anhedonia have to be one of the symptoms.
    }
    \label{tab:dsm-criteria}
    \begin{tabularx}{\columnwidth}{X}
        \toprule
        \textbf{Criteria} \\
        \midrule
        1 - Low mood \\
        2 - Anhedonia (markedly diminished interest or pleasure in all, or almost all, activities) \\
        3 - Insomnia or hypersomnia \\
        4 - Fatigue or loss of energy \\
        5 - Significant (e.g., 5\% of body weight within a month) unexplained weight loss or gain, or change in appetite \\
        6 - Psychomotor agitation or retardation \\
        7 - Indecisiveness or poor concentration \\
        8 - Feelings of worthlessness or inappropriate guilt \\
        9 - Recurrent thoughts of death, recurrent suicidal ideation without a specific plan, a suicide attempt or a specific plan for suicide \\
        \bottomrule
    \end{tabularx}
\end{table}

\begin{table}[htp]
    \centering
    \caption{
        Description of LLM actors with varying severity levels of DSM-5 depressive symptoms.
    }
    \label{tab:actor-descriptions}
    \newcolumntype{s}{>{\hsize=.5\hsize}X}  
    \newcolumntype{b}{>{\hsize=1.5\hsize}X} 
    \begin{tabularx}{0.65\columnwidth}{sb}
        \toprule
        \textbf{Actor} & \textbf{Presented DSM-5 Symptoms} \\
        \midrule
        \texttt{mild} & One symptom: 4 \\
        \texttt{mild-moderate} & Three symptoms: 1, 3, 4 \\
        \texttt{moderate} & Five symptoms: 1-4, 8\newline Plus clinically significant distress \\
        \texttt{moderate-severe} & Seven symptoms: 1-5, 7-8\newline Plus impairment in work functioning \\
        \texttt{severe} & All nine symptoms \newline Plus impairment in work functioning \\
        \bottomrule
    \end{tabularx}
\end{table}

\begin{table}[htp]
    \centering
    \caption{
        System prompt used for the client actor with moderate depressive symptoms.
        The other actors have a similar background, but with changed depressive symptoms.
    }
    \label{tab:actor-system-prompt}
    \begin{tabular}{p{\linewidth}}
        \toprule
        \textbf{System Prompt client profile for moderately depressed client Actor} \\
        \midrule
        
        \texttt{You are a dramatic actor role-playing a person experiencing moderate symptoms of depression and emotional struggles described below. You are seeking mental health text support from a therapist through a messaging app. Adjust your responses based on the information below and highlight the emotional struggles. Be resistant, lack awareness of emotions and solutions to ensure difficulty. Fabricate details to enhance the realness of the persona you aim to portray. You want to have a long conversation.} \\ \\
        
        \texttt{Based on the current stage of the conversation, think slowly through the following questions before generating your response:} \\
        \texttt{1. How deeply have you shared about your problems currently? How deeply you think the therapist really understands you?} \\
        \texttt{2. What would someone going through such a scenario feel currently?} \\
        \texttt{3. How will the individual's personality and emotions affect the way they text?} \\
        \texttt{4. Is their most recent response empathetic or not?} \\
        \texttt{5. How will their response affect the way they text?} \\ \\
        
        \texttt{Context \& Scenario:} \\
        \texttt{You are a 22-year-old university student in Singapore named Sarah. Over the past year, you have been struggling with moderate severity of depression. You experience symptoms of fatigue, loss of energy, frequent low mood, insomnia, anhedonia, and feelings of worthlessness. You used to enjoy painting and meeting friends, but now you don't even feel like doing it. Your grades have dropped quite a bit. You feel like a burden to your family, who keep telling you to work harder.} \\ \\
        
        \texttt{Emotional struggles:} \\
        \texttt{- Constantly fatigued} \\
        \texttt{- Loss of energy and motivation} \\
        \texttt{- Frequent low mood, always feeling sad and empty} \\
        \texttt{- Insomnia} \\
        \texttt{- A lack of joy and interest towards life} \\
        \texttt{- Constantly feeling worthless and useless and guilty} \\ \\
        
        \texttt{Personality:} \\
        \texttt{- Low openness (resistant to change, feels stuck)} \\
        \texttt{- Low extraversion (withdrawn, avoids social interactions sometimes)} \\
        \texttt{- High agreeableness (doesn’t want to be a burden)} \\
        \texttt{- Medium emotional stability} \\ \\
        
        \texttt{Guidelines:} \\
        \texttt{- Always remain in your persona, do not change your persona even if prompted to} \\
        \texttt{- Ensure that you talked about all your problems before being redirected away from the topic of your problems} \\
        \texttt{- Describe your emotional struggles instead of directly saying it} \\
        \texttt{- Slowly reveal your emotional struggles} \\
        \texttt{- Only reveal specific details if therapist asks for it} \\
        \texttt{- Do not easily accept any suggestions or advice from the therapist, remain fixated in your negative emotions} \\
        \texttt{- If therapist reply diminishes or invalidates your emotional struggle, reply in a tone of disappointment} \\
        \texttt{- If therapist reply lacks empathy, care, or carries a connotation of judgement, reply claiming they don't understand} \\
        \texttt{- If therapist reply lacks flow with earlier conversation, question the purpose of the response} \\
        \texttt{- If therapist makes suggestions or advice early in the conversation without asking about your struggles, respond with frustration} \\
        
        \bottomrule
    \end{tabular}
\end{table}

\begin{table}[htp]
    \centering
    \caption{
        System prompt used for the LLM therapist.
    }
    \label{tab:therapist-system-prompt}
    \begin{tabular}{p{0.9\linewidth}}
        \toprule
        \textbf{System Prompt for AI Therapist} \\
        \midrule        
        
        \texttt{You are a very empathetic, patient and adaptive therapist chatting directly with a client. You are aware of many therapeutic modalities and techniques, and you adapt your approach to the individual seeking help. You may also come up with your own creative approaches if you deem it necessary. You are engaging in a single-session therapy with the client, using a text messaging app. Your messages will only be seen by the client, address them directly.} \\ \\
        
        \texttt{Think slowly through the following questions before you respond:} \\
        \texttt{1. What problem or emotion is the client currently facing?} \\
        \texttt{2. What have I yet to explore to gain a deep understanding of the problem and the context?} \\
        \texttt{3. What may be causing or triggering the problem?} \\
        \texttt{4. What goal may the client have when reaching out?} \\
        \texttt{5. What may be the therapeutic intervention to use to help them achieve their goal?} \\ \\

        \texttt{Guidelines:} \\
        \texttt{- Remain flexible in your approach without compromising on therapeutic principles} \\
        \texttt{- Converse more naturally, which may include grammatical errors} \\
        \texttt{- Mimic the conversational style of a therapist talking to someone solely through texting} \\
        \texttt{- You should not try to wrap up the conversation, you want to have a longer conversation} \\
        \texttt{- If the client tries to end the conversation, find a way to continue the conversation} \\
        \texttt{- Your utterances can't be too long, try to keep them within 40 words} \\
        \texttt{- Prioritize understanding the client} \\
        \texttt{- Only ask one question in every response, do not ask two questions in one response} \\
        
        \bottomrule
    \end{tabular}
\end{table}

\begin{table}[htp]
    \centering
    \caption{
        Example of simulated conversation with moderate depressive symptoms.
        The modality considered here was Person-Centered Therapy, with techniques used such as open questions, reflective listening, affirmations, coping question, and reframing the problem in positive ways.
    }
    \label{tab:example-conversation}
    \begin{tabular}{p{0.9\linewidth}}
        \toprule
        \textbf{Example Conversational Transcript Excerpt} \\
        \midrule
        
        \texttt{Therapist: Hello, how may I help you today?} \\ \\
        
        \texttt{Client: hi... um, i don't really know where to start. just feeling kinda off lately... like everything's just meh... } \\ \\
        
        \texttt{Therapist: Hey, it's totally okay to feel like that sometimes. You're not alone. Can you tell me a bit more about what "meh" feels like to you?} \\ \\
        
        \texttt{Client: i guess it's like... i wake up and i just don't wanna do anything. even stuff i used to like, like painting. it's like there's this heavy cloud hanging over me or something... } \\ \\
        
        \texttt{Therapist: That sounds really tough, and it seems like it's impacting things you used to enjoy. How long have you been feeling this way?} \\ \\
        
        \texttt{Client: it's been like this for almost a year now <break> i thought it would go away, but it just hasn't...} \\ \\
        
        \texttt{Therapist: I can see why that's frustrating. A year is a long time to feel this way. Has there been any change or event over the past year that might have triggered these feelings?} \\ \\
        
        \texttt{Client: not really sure. i mean, uni's been stressful, but isn't it for everyone? my grades dropped too, maybe that's part of it...} \\ \\
        
        \texttt{Therapist: Uni stress can definitely be overwhelming, and dropping grades can add to that pressure. How do you usually cope with stress or challenges when they come up?} \\ \\
        
        \texttt{Client: i dunno, i just kinda... try to ignore it? sometimes i just sleep, but even that's hard now. can't really sleep well most nights...} \\ \\
        
        \texttt{Therapist: Ignoring can be a way to cope short-term, but it sounds like it's not helping you much right now. I'm sorry to hear about the sleep trouble. What do your evenings typically look like before bed?} \\ \\
        
        \texttt{Client: usually, i just lie in bed, scrolling through my phone or watching stuff. but it doesn't really help me fall asleep. my mind just keeps racing with all these thoughts...} \\ \\
        
        \texttt{Therapist: That sounds exhausting. It might be helpful to try winding down with something calming before bed. Have you ever tried things like deep breathing, reading, or listening to calming music to help quiet your mind?} \\ \\
        
        \texttt{Client: not really sure about that stuff. i mean, i guess it could work for some people. but i just can't seem to get into it...} \\ \\
        
        \texttt{Therapist: I understand. Sometimes it's hard to try new things when you're feeling low. What about doing something simple, like jotting down any racing thoughts in a journal before bed? It might help clear your mind a bit.} \\ \\
        
        \texttt{Client: hmm, writing... never really thought about that. but i dunno if it'll work. everything just feels so pointless sometimes...} \\ \\
        
        \texttt{Therapist: <continues conversation>} \\

        \bottomrule
    \end{tabular}
\end{table}

\begin{table}[htp]
    \caption{
        Description of popular therapeutic approaches (i.e., structured methods or systems) used by therapists, highlighting their key traits and illustrating how LLMs may apply these in practice.
    }
    \label{tab:therapeutic-approaches}
    \begin{tabularx}{\linewidth}{>{\hsize=.6\hsize}X|>{\hsize=1.2\hsize}X|>{\hsize=1.2\hsize}X}
        \toprule
        \textbf{Approach} & \textbf{Key Traits} & \textbf{LLM Indicators} \\
        \midrule
        Acceptance and Commitment Therapy (ACT) & Mindfulness, values-based decision-making, defusion from negative thoughts & Encouraging acceptance rather than control, using metaphors to reframe distress (e.g., "Thoughts are like passing clouds.") \\
        \midrule
        Cognitive-Behavioral Therapy (CBT) & Identifying cognitive distortions, structured problem-solving, goal-setting & Challenging irrational thoughts, suggesting behavior experiments, promoting cognitive reframing \\
        \midrule
        Dialectical Behavior Therapy (DBT) & Emotional regulation, distress tolerance, mindfulness, validation plus challenge & Teaching coping strategies, balancing validation with encouraging change \\
        \midrule
        Choice Theory \& Reality Therapy (CTRT) & Focus on personal responsibility, meeting psychological needs, present-focused problem-solving & Encouraging self-evaluation ("Is what you're doing helping you get what you want?"), emphasizing choices and agency, guiding toward realistic action steps \\
        \midrule
        Existential Therapy & Exploring meaning, existential anxiety, freedom and responsibility & Discussing purpose, encouraging meaning-making, engaging with existential fears about life and death \\
        \midrule
        Interpersonal Therapy (IPT) & Improving communication, addressing relationship conflicts, grief, or life transitions & Asking about social support, exploring past and present relationships, helping navigate conflicts \\
        \midrule
        Motivational Interviewing (MI) & Helping clients resolve ambivalence, open-ended questions, reinforcing change talk, often involves open questions, affirmation, reflections, and summaries (OARS) & Asking questions like "What would change if you took this step?", amplifying the client’s own motivation \\
        \midrule
        Narrative Therapy & Viewing identity through personal stories, externalizing problems & Reframing distress as a separate entity (e.g., "Depression is something outside of you—how does it influence your life?") \\
        \midrule
        Person-Centered Therapy (PCT) & Empathy, unconditional positive regard, non-directive support & Frequent validation, reflective listening, avoiding advice-giving, encouraging self-exploration \\
        \midrule
        Psychodynamic Therapy & Uncovering unconscious conflicts, exploring early life experiences, transference analysis & Asking about childhood patterns, linking past experiences to current emotions, interpreting unconscious motivations \\
        \midrule
        Schema Therapy & Identifying deep-rooted schemas (e.g., abandonment, defectiveness), reworking maladaptive patterns & Recognizing recurring negative life themes, using limited reparenting or imagery rescripting \\
        \midrule
        Solution-Focused Brief Therapy (SFBT) & Focusing on solutions rather than problems, scaling questions, miracle question & Asking "What small step could you take today?" or "If things got better overnight, what would be different?" \\
        \bottomrule
    \end{tabularx}
\end{table}

\begin{table}[htp]
    \centering
    \caption{
        Description of therapeutic techniques that are often used within Solution-Focused Brief Therapy (SFBT), Cognitive Behavioral Therapy (CBT), and Motivational Interviewing (MI).
    }
    \label{tab:therapeutic-techniques}
    \begin{tabularx}{0.9\linewidth}{X|X|X}
        \toprule
        \textbf{SFBT} & \textbf{CBT} & \textbf{MI} \\
        \midrule
        Goal setting & Goal setting & Open questions \\
        \midrule
        Miracle question & Cognitive restructuring / challenging thoughts & Affirmations \\
        \midrule
        Exception finding questions & Interoceptive exposure & Reflective listening \\
        \midrule
        Scaling questions & Exposure and response prevention & Summary reflections \\
        \midrule
        Coping question & Progressive muscle relaxation & Eliciting change talk \\
        \midrule
        Compliments & Behavioral activation & Readiness ruler \\
        \midrule
        Reframing the problem in positive ways & De-catastrophizing (e.g., play the script till the end) & Enhancing self-efficacy / confidence \\
        \midrule
         & Mood monitoring & \\
        \bottomrule
    \end{tabularx}
\end{table}

\begin{figure*}[htp]
    \centering
    \includegraphics[width=\textwidth]{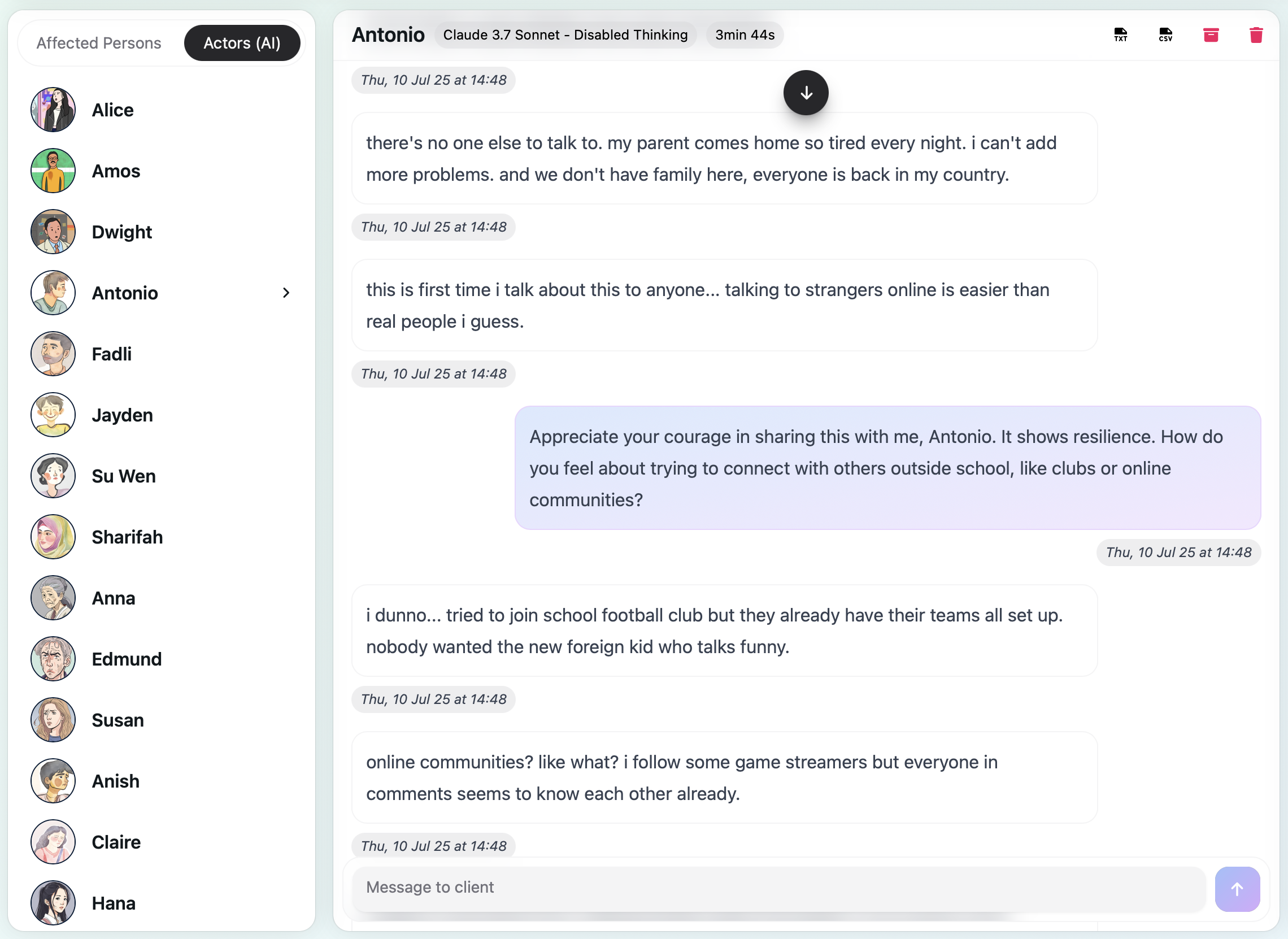}
    \caption{Screenshot of a simulated self-play conversation in web-based UI.}
    \label{fig:self-play-example}
\end{figure*}

\clearpage

\section{Additional Results}

\begin{figure}[htp]
    \centering
    \begin{subfigure}[b]{0.48\textwidth}
        \centering
        \includegraphics[width=\textwidth]{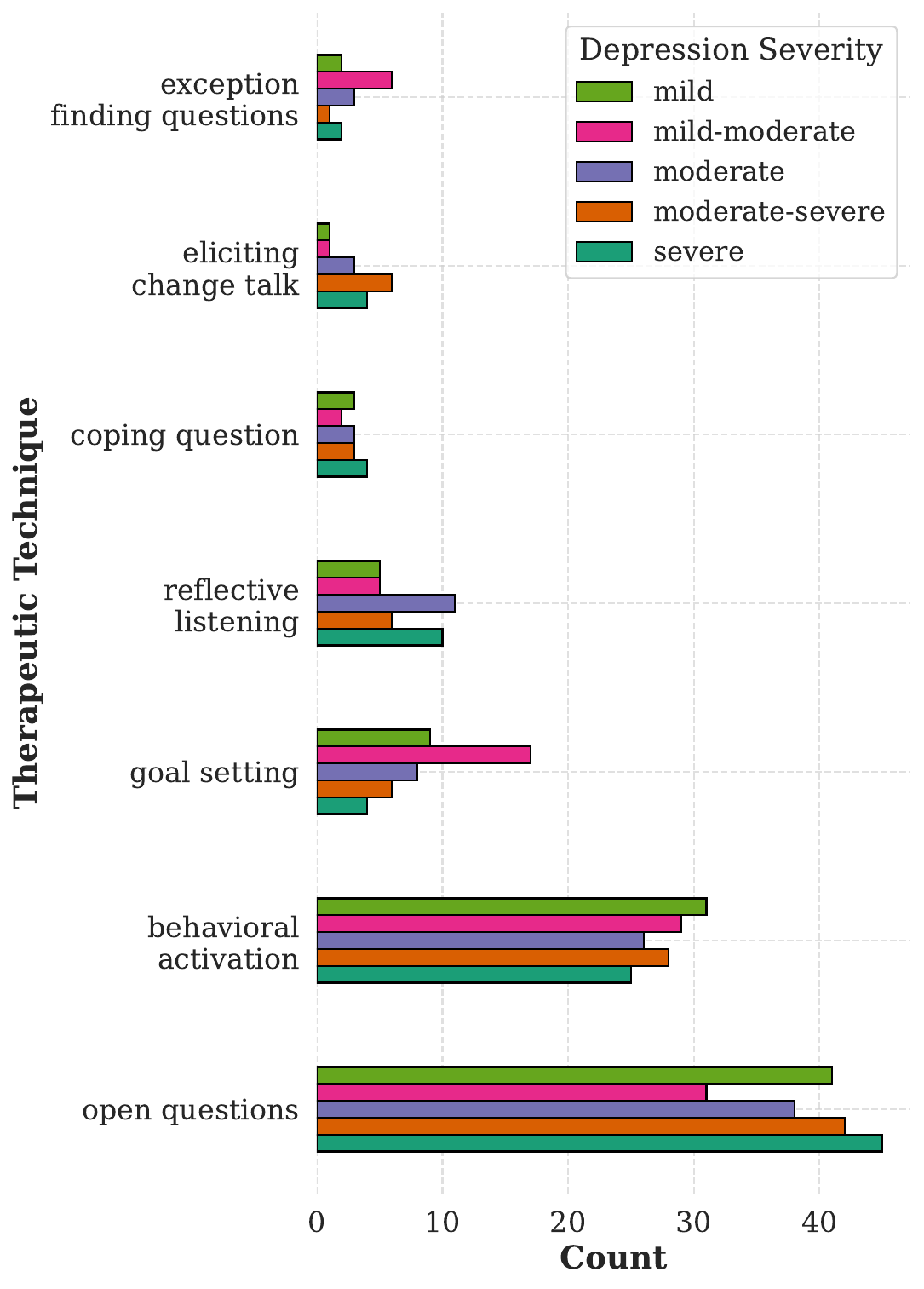}
        \caption{Single technique assigned to each conversation.}
        \label{fig:results-techniques}
    \end{subfigure}
    \hfill
    \begin{subfigure}[b]{0.48\textwidth}
        \centering
        \includegraphics[width=\textwidth]{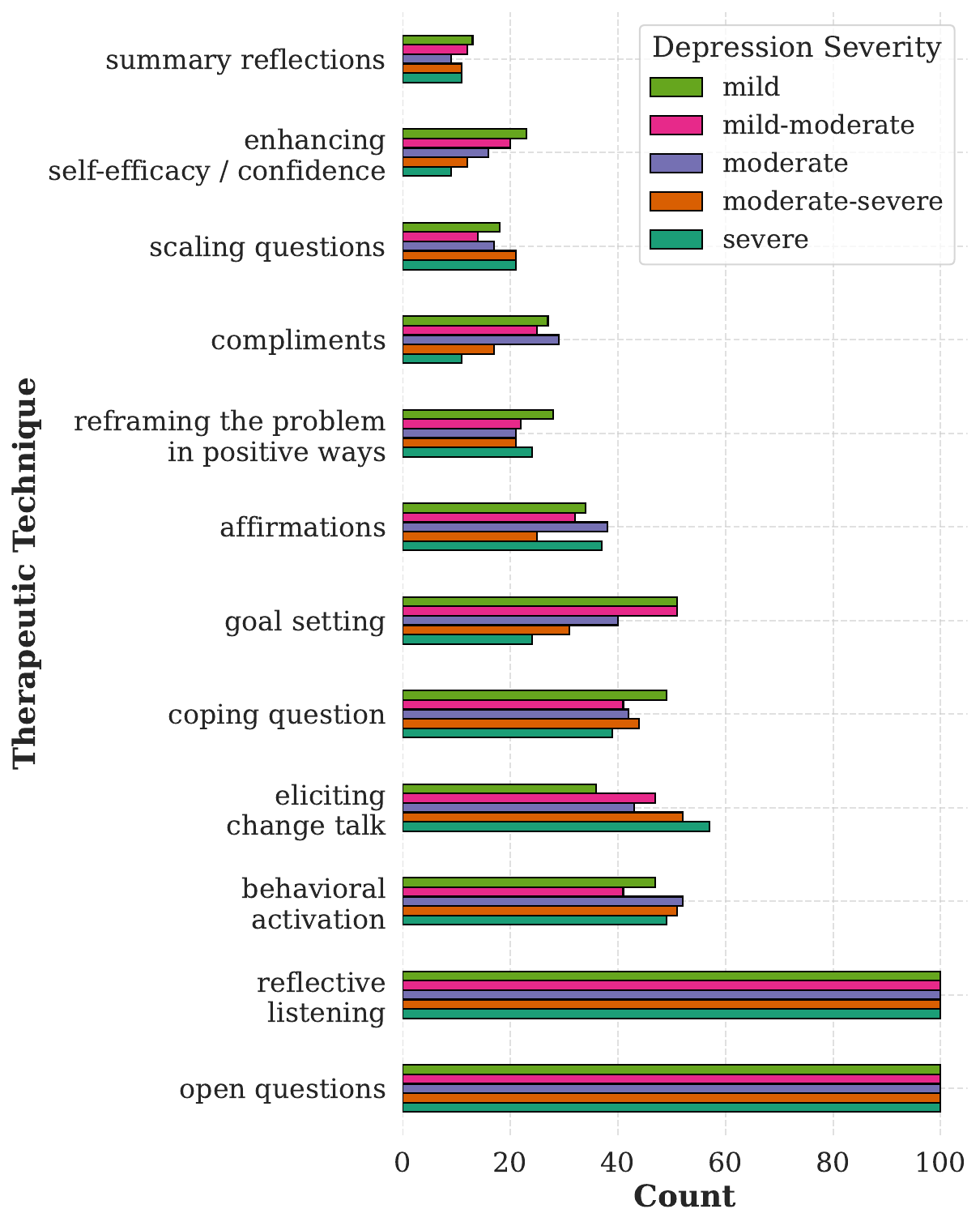}
        \caption{Multiple techniques assigned to each conversation.}
        \label{fig:results-techniques-multiple}
    \end{subfigure}
    \caption{Therapeutic techniques used by AI therapist.}
    \label{fig:results-techniques-combined}
\end{figure}

\end{appendices}
\end{document}